\documentclass[sigplan,10pt,dvipsnames]{acmart}
\usepackage{amscd,amsmath, cite,multirow}
\usepackage[justification=centering,skip=0pt,belowskip=-15pt,aboveskip=0pt]{caption}
\usepackage{enumitem}
\usepackage{subcaption}
\usepackage{mdframed}

\usepackage{comment}

\newcommand\remove[1]{}
\makeatother

\usepackage{tikz}

\newcommand{\hlcolor}{Yellow!35}
\newcommand{\hlcolorTwo}{LimeGreen!35}
\makeatletter
\newenvironment{btHighlight}[1][]
{\begingroup\tikzset{bt@Highlight@par/.style={#1}}\begin{lrbox}{\@tempboxa}}
{\end{lrbox}\bt@HL@box[bt@Highlight@par]{\@tempboxa}\endgroup}

\newcommand\btHL[1][]{%
  \begin{btHighlight}[#1]\bgroup\aftergroup\bt@HL@endenv%
}
\def\bt@HL@endenv{%
  \end{btHighlight}%
  \egroup
}
\newcommand{\bt@HL@box}[2][]{%
  \tikz[#1]{%
    \pgfpathrectangle{\pgfpoint{1pt}{0pt}}{\pgfpoint{\wd #2}{\ht #2}}%
    \pgfusepath{use as bounding box}%
    \node[anchor=base west, fill=\hlcolor,outer sep=0pt,inner xsep=1pt, inner ysep=0pt, rounded corners=2pt, minimum height=\ht\strutbox+2pt,#1]{\raisebox{1pt}{\strut}\strut\usebox{#2}};
  }%
}

\newenvironment{btHighlightTwo}[1][]
{\begingroup\tikzset{bt@HighlightTwo@par/.style={#1}}\begin{lrbox}{\@tempboxa}}
{\end{lrbox}\bt@HLTwo@box[bt@HighlightTwo@par]{\@tempboxa}\endgroup}

\newcommand\btHLTwo[1][]{%
  \begin{btHighlightTwo}[#1]\bgroup\aftergroup\bt@HLTwo@endenv%
}
\def\bt@HLTwo@endenv{%
  \end{btHighlightTwo}%
  \egroup
}
\newcommand{\bt@HLTwo@box}[2][]{%
  \tikz[#1]{%
    \pgfpathrectangle{\pgfpoint{1pt}{0pt}}{\pgfpoint{\wd #2}{\ht #2}}%
    \pgfusepath{use as bounding box}%
    \node[anchor=base west, fill=\hlcolorTwo,outer sep=0pt,inner xsep=1pt, inner ysep=0pt, rounded corners=2pt, minimum height=\ht\strutbox+2pt,#1]{\raisebox{1pt}{\strut}\strut\usebox{#2}};
  }%
}

\usepackage{listings}
\usepackage{etoolbox}
\newtoggle{InString}{}
\togglefalse{InString}
\newcommand*{\ColorIfNotInString}[1]{\iftoggle{InString}{#1}{\color{blue}#1}}%
\newcommand*{\ProcessQuote}[1]{#1\iftoggle{InString}{\global\togglefalse{InString}}{\global\toggletrue{InString}}}%
\definecolor{code_indent}{HTML}{CCCCCC}

\usepackage{multicol}

\acmConference[PL'18]{ACM SIGPLAN Conference on Programming Languages}{January 01--03, 2018}{New York, NY, USA}
\acmYear{2018}
\acmISBN{} 
\acmDOI{} 
\startPage{1}
\setcopyright{none}
\begin{document}

\title{Peformance Prediction for Coarse-Grained Locking: MCS Case}

\author{Vitaly Aksenov}
\affiliation{ITMO University}

\author{Daniil Bolotov}
\affiliation{ITMO University}

\author{Petr Kuznetsov}
\affiliation{T\'el\'ecom Paris}

\date{August 2021}

\remove{
\begin{abstract}
A standard design pattern found in many concurrent data
structures, such as hash tables or ordered containers,
is alternation of \emph{parallelizable} sections that incur no data
conflicts and \emph{critical} sections that must run sequentially and
are protected with locks.    
It was already shown that simple stochastic analysis can predict the throughput of coarse-grained lock-based algorithms using CLH lock. 
In this short paper, we extend this analysis to algorithms based on the popular MCS lock.
\end{abstract}
}

\maketitle

\section{Introduction}
Analytical prediction of the performance of a concurrent data structure is a challenging task.
Earlier works proposed prediction frameworks for  
lock-free data structures~\citep{tsigas15, tsigas16, tsigas18} and lock-based ones~\citep{AAK18-ba} in which the coarse-grained critical section is protected with CLH lock~\citep{craig1993building}. 
In this paper, we make the next step and describe an analytical prediction framework for concurrent data structure based on MCS lock~\citep{mellor1991algorithms}. 
The MCS framework turned out to be more elaborated than the CLH one.    

\section{Model}
Consider a concurrent system with $N$ processes that obey  
the following simple \emph{uniform} scheduler:
in every time tick, each process performs a step of computation.
This scheduler,  resembling the well-known PRAM model~\citep{jaja1992introduction}, 
appears to be a reasonable approximation of real-life concurrent systems.
Suppose 
that the processes share a data structure exporting a single abstract
\texttt{operation}$()$.
Assume that one operation induces work of size $P$ and incurs no synchronization, and one process performs work $\alpha$ in a time unit. 
Then the resulting throughput is $N \cdot \alpha / P$ operations per time unit.
%
One way to evaluate the constant $\alpha$ is to experimentally count
the total number $F$ of operations, each of work $P$,   
completed by $N$ processes in time $T$.
Then we get $\alpha=PF/(TN)$.
The longer is $T$, the more  accurate is the estimation of $\alpha$.

Consider a simple lock-based concurrent data structure, where   
\texttt{operation()} is implemented using the pseudocode in Figure~\ref{fig:operation}.

\begin{figure}
\begin{lstlisting}
operation():
  lock.lock() 
  for i in 1..C:  |\label{line:crit1}|    
    nop |\label{line:crit2}|
  lock.unlock()
  for i in 1..P: |\label{line:par1}|
    nop |\label{line:par2}|
\end{lstlisting}
\caption{The coarse-grained operation}
\label{fig:operation}
\end{figure}


The abstract machine used in our 
analytical throughput prediction obey the following conditions.

First, we assume that the coherence of caches is maintained by
a variant of MESI protocol~\citep{papamarcos1984low}.
%
%
Each cache line can be in one of four states: \texttt{Modified} (\texttt{M}),
\texttt{Exclusive} (\texttt{E}), \texttt{Shared} (\texttt{S}) and \texttt{Invalid} (\texttt{I}).                        
MESI regulates transitions between states of a cache line
depending on the request (read or write) to the cache line by a process
or on the request to the memory bus.
The important transitions for us are:
(1)~upon reading, the state of the cache line changes from any state to \texttt{S},
and, if the state was \texttt{I}, then a \emph{read request} is sent to the bus;
(2)~upon writing, the state of the cache line becomes \texttt{M},
and, if the state was \texttt{S} or \texttt{I}, an \emph{invalidation request} is sent to the bus. 

We assume that the caches are \emph{symmetric}:
for each MESI state $st$, there exist two constants $R_{st}$ and $W_{st}$
such that any read from any cache line with status $st$ takes $R_{st}$ work and
any write to a cache line with status $st$ takes $W_{st}$ work. 
%
%
David et al.~\citep{david2013everything} showed that for an Intel Xeon
machine (similar to the one we use in our experimental validation
below), 
given the relative location of a cache line with respect to the process
(whether they are located on the same socket or not), the
following hypotheses hold:
(1)~writes induce the same work, regardless of the state of the cache line;
(2)~\texttt{swap}s, not concurrent with other \texttt{swap}s, induce the same work as writes;
(3)~reads from the invalid cache line costs more than a write.
Therefore, we assume 
that (1)~$W = W_M = W_E = W_S = W_I$;
(2)~any contention-free \texttt{swap} induces a work of size $W$; and (3)~$R_I \geq W$.

\section{MCS lock}
As the first step, we replace \texttt{lock.lock()} and \texttt{lock.unlock()} in Figure~\ref{fig:operation} with MCS implementation: the resulting code is  presented in Figure~\ref{fig:coarse}.

We can then calculate the cost of each line: Lines~\ref{line:setnext}~and~\ref{line:setlocked} cost $W$ each; Line~\ref{line:swap} costs $W$ in the uncontended case and $X$ in the contended one; if the lock is taken, i.e., a condition in Line~\ref{line:pred:condition} returns \textit{true}, Line~\ref{line:queue} costs $W$ and Line~\ref{line:wait-critical} costs $R_I$; the critical section costs $C$; if the lock is free, then the read in Line~\ref{line:release-1} costs $R_M$, otherwise, it costs $R_I$; Line~\ref{line:cas} costs $W$ in the uncontended case and $X$ in the contended one; Line~\ref{line:unlocked-next} costs $W$; and, finally, the parallel section costs $P$.

In our theoretical analysis, we distinguish two cases: (1)~when the lock is always free, i.e., \texttt{swap} in Line~\ref{line:swap} returns \texttt{null}, the conditional statement in Line~\ref{line:pred:condition} does not succeed, and the conditional statement in Line~\ref{line:release-1} and CAS in Line~\ref{line:cas} always succeeds; and (2)~when the lock is always taken, i.e., the conditional statement in Line~\ref{line:pred:condition} succeeds and the conditional statement in Line~\ref{line:release-1} does not succeed.

\begin{figure*}
  \begin{minipage}{0.79\textwidth}
    \begin{subfigure}{\linewidth}
      \resizebox{\textwidth}{!}{
        \begin{tikzpicture}

\node at (-4.8,0.6) {$1$};
\draw (-4.5,0.6) -- (-4.3,0.6);
\draw (-2.4,0.8) -- (-4.3,0.8) -- (-4.3,0.4) -- (-2.4,0.4);
\node at (-2.75,0.6) {$X$};
\node at (-3.4,0.6) {$W$};
\node at (-4,0.6) {$W$};
\draw (-3.1, 0.8) -- (-3.1, 0.4);
\draw (-3.7, 0.8) -- (-3.7, 0.4);

\draw[fill=blue!10] (-2.4,0.8) -- (-2.4,0.4) -- (0.4,0.4) -- (0.4,0.8) -- cycle;
\draw (-1,0.8) -- (-1,0.4);
\draw (-0.2,0.8) -- (-0.2,0.4);
\node at (-1.7,0.6) {$C$};
\node at (-0.6,0.6) {$R_I$};
\node at (0.1,0.6) {$W$};
\draw [fill=red!10] (10.2,0.4) -- (0.4,0.4) -- (0.4,0.8) -- (10.2,0.8) -- cycle;
\draw (8.4,0.8) -- (8.4,0.4);
\draw (9,0.8) -- (9,0.4);
\draw (9.6,0.8) -- (9.6,0.4);
\node at (4.4,0.6) {$P$};
\node at (8.7,0.6) {$W$};
\node at (9.3,0.6) {$W$};
\node at (9.9,0.6) {$W$};
\draw [fill=blue!10] (10.2,0.8) -- (10.2,0.4) -- (12.7,0.4) -- (12.7,0.8) -- cycle;
\draw (11.6,0.8) -- (11.6,0.4);
\draw (12.1,0.8) -- (12.1,0.4);
\draw (12.7,0.8) -- (12.7,0.4);
\node at (10.9,0.6) {$C$};
\node at (11.85,0.6) {$R_M$};
\node at (12.4,0.6) {$W$};
\draw [fill=red!10] (14,0.8) -- (12.7,0.8) -- (12.7,0.4) -- (14,0.4);

\node at (-4.8,0) {$2$};
\draw (-4.5,0) -- (-4.3,0);
\draw (-3.1,0.2) -- (-4.3,0.2) -- (-4.3,-0.2) -- (-3.1,-0.2) --cycle;
\node at (-3.4,0) {$W$};
\node at (-4,0) {$W$};
\draw (-3.7, 0.2) -- (-3.7, -0.2);
\draw (-3.1, 0) -- (-2.4, 0);
\draw (-2.4,-0.2) -- (-1.1,-0.2) -- (-1.1,0.2) -- (-2.4,0.2) -- cycle;
\draw (-1.7,-0.2) -- (-1.7,0.2);
\node at (-2.05,0) {$X$};
\node at (-1.4,0) {$W$};
\draw (-1.1,0) -- (0.8,0);
\draw [fill=blue!10] (0.4,0.2) -- (0.4,-0.2) -- (4,-0.2) -- (4,0.2) -- cycle;
\draw (1.2,0.2) -- (1.2,-0.2);
\draw (2.6,0.2) -- (2.6,-0.2);
\draw (3.4,0.2) -- (3.4,-0.2);
\node at (0.8,0) {$R_I$};
\node at (1.9,0) {$C$};
\node at (3,0) {$R_I$};
\node at (3.7,0) {$W$};
\draw [fill=red!10] (13.8,0.2) -- (4,0.2) -- (4,-0.2) -- (13.8,-0.2) -- cycle;
\draw (12,0.2) -- (12,-0.2);
\draw (12.6,0.2) -- (12.6,-0.2);
\draw (13.2,0.2) -- (13.2,-0.2);
\node at (8,0) {$P$};
\node at (12.3,0) {$W$};
\node at (12.9,0) {$W$};
\node at (13.5,0) {$W$};
\draw [fill=blue!10] (14,0.2) -- (13.8,0.2) -- (13.8,-0.2) -- (14,-0.2);

\node at (-4.8,-0.4) {$\vdots$};

\node at (-4.8,-1.2) {$N$};
\draw (-4.5,-1.2) -- (-4.3,-1.2);
\draw (-3.1,-1) -- (-4.3,-1) -- (-4.3,-1.4) -- (-3.1,-1.4) --cycle;
\node at (-3.4,-1.2) {$W$};
\node at (-4,-1.2) {$W$};
\draw (-3.7, -1) -- (-3.7, -1.4);
\draw (-3.1,-1.2) -- (-0.8,-1.2);
\draw (-0.8,-1) -- (-0.8,-1.4) -- (0.5,-1.4) -- (0.5,-1) -- cycle;
\node at (-0.45,-1.2) {$X$};
\node at (0.2,-1.2) {$W$};
\draw (-0.1,-1.4) -- (-0.1,-1);
\draw (0.5,-1.2) -- (7.2,-1.2);
\draw [fill=blue!10] (6.2,-1) -- (6.2,-1.4) -- (9.5,-1.4) -- (9.5,-1) -- cycle;
\draw (7,-1) -- (7,-1.4);
\draw (8.4,-1) -- (8.4,-1.4);
\draw (8.9,-1) -- (8.9,-1.4);
\node at (6.6, -1.2) {$R_I$};
\node at (7.7,-1.2) {$C$};
\node at (8.65,-1.2) {$R_M$};
\node at (9.2,-1.2) {$W$};
\draw [fill=red!10] (14,-1) -- (9.5,-1) -- (9.5,-1.4) -- (14,-1.4);

\draw [dashed,line width=2] (9.5,1.2) -- (9.5,-1.8);

\node at (15.8,0) {};

\end{tikzpicture}
      }
      \captionsetup{width=.8\linewidth}%
      \caption{Case 1: $P + 2W \geq (N - 1) \cdot (C + 2R_I + W) + R_I - R_M$.
         The lock is always free.}
      \label{fig:schedule-1}
    \end{subfigure}
    \begin{subfigure}{\linewidth}
      \resizebox{\textwidth}{!}{
        \begin{tikzpicture}

\node at (-4.8,0.6) {$1$};
\draw (-4.5,0.6) -- (-4.3,0.6);
\draw (-2.4,0.8) -- (-4.3,0.8) -- (-4.3,0.4) -- (-2.4,0.4);
\node at (-2.75,0.6) {$X$};
\node at (-3.4,0.6) {$W$};
\node at (-4,0.6) {$W$};
\draw (-3.1, 0.8) -- (-3.1, 0.4);
\draw (-3.7, 0.8) -- (-3.7, 0.4);

\draw[fill=blue!10] (-2.4,0.8) -- (-2.4,0.4) -- (0.4,0.4) -- (0.4,0.8) -- cycle;
\draw (-1,0.8) -- (-1,0.4);
\draw (-0.2,0.8) -- (-0.2,0.4);
\node at (-1.7,0.6) {$C$};
\node at (-0.6,0.6) {$R_I$};
\node at (0.1,0.6) {$W$};
\draw [fill=red!10] (6.8,0.4) -- (0.4,0.4) -- (0.4,0.8) -- (6.8,0.8) -- cycle;
\draw (4.4,0.8) -- (4.4,0.4);
\draw (5,0.8) -- (5,0.4);
\draw (5.6,0.8) -- (5.6,0.4);
\draw (6.2,0.8) -- (6.2,0.4);
\node at (2.4,0.6) {$P$};
\node at (4.7,0.6) {$W$};
\node at (5.3,0.6) {$W$};
\node at (5.9,0.6) {$W$};
\node at (6.5,0.6) {$W$};
\draw (6.8,0.6) -- (9.7,0.6);
\draw [fill=blue!10] (9.8,0.8) -- (9.8,0.4) -- (13.4,0.4) -- (13.4,0.8) -- cycle;
\draw (10.6,0.8) -- (10.6,0.4);
\draw (12,0.8) -- (12,0.4);
\draw (12.8,0.8) -- (12.8,0.4);
\node at (10.2,0.6) {$R_I$};
\node at (11.3,0.6) {$C$};
\node at (12.4,0.6) {$R_I$};
\node at (13.1,0.6) {$W$};
\draw [fill=red!10] (14,0.8) -- (13.4,0.8) -- (13.4,0.4) -- (14,0.4);

\node at (-4.8,0) {$2$};
\draw (-4.5,0) -- (-4.3,0);
\draw (-3.1,0.2) -- (-4.3,0.2) -- (-4.3,-0.2) -- (-3.1,-0.2) --cycle;
\node at (-3.4,0) {$W$};
\node at (-4,0) {$W$};
\draw (-3.7, 0.2) -- (-3.7, -0.2);
\draw (-3.1, 0) -- (-2.4, 0);
\draw (-2.4,-0.2) -- (-1.1,-0.2) -- (-1.1,0.2) -- (-2.4,0.2) -- cycle;
\draw (-1.7,-0.2) -- (-1.7,0.2);
\node at (-2.05,0) {$X$};
\node at (-1.4,0) {$W$};
\draw (-1.1,0) -- (0.8,0);
\draw [fill=blue!10] (0.4,0.2) -- (0.4,-0.2) -- (4,-0.2) -- (4,0.2) -- cycle;
\draw (1.2,0.2) -- (1.2,-0.2);
\draw (2.6,0.2) -- (2.6,-0.2);
\draw (3.4,0.2) -- (3.4,-0.2);
\node at (0.8,0) {$R_I$};
\node at (1.9,0) {$C$};
\node at (3,0) {$R_I$};
\node at (3.7,0) {$W$};
\draw [fill=red!10] (10.4,0.2) -- (4,0.2) -- (4,-0.2) -- (10.4,-0.2) -- cycle;
\draw (8,0.2) -- (8,-0.2);
\draw (8.6,0.2) -- (8.6,-0.2);
\draw (9.2,0.2) -- (9.2,-0.2);
\draw (9.8,0.2) -- (9.8,-0.2);
\node at (6,0) {$P$};
\node at (8.3,0) {$W$};
\node at (8.9,0) {$W$};
\node at (9.5,0) {$W$};
\node at (10.1,0) {$W$};
\draw (10.4,0) -- (13.6,0);
\draw [fill=blue!10] (14,0.2) -- (13.3,0.2) -- (13.3,-0.2) -- (14,-0.2);

\node at (-4.8,-0.4) {$\vdots$};

\node at (-4.8,-1.2) {$N$};
\draw (-4.5,-1.2) -- (-4.3,-1.2);
\draw (-3.1,-1) -- (-4.3,-1) -- (-4.3,-1.4) -- (-3.1,-1.4) --cycle;
\node at (-3.4,-1.2) {$W$};
\node at (-4,-1.2) {$W$};
\draw (-3.7, -1) -- (-3.7, -1.4);
\draw (-3.1,-1.2) -- (-0.8,-1.2);
\draw (-0.8,-1) -- (-0.8,-1.4) -- (0.5,-1.4) -- (0.5,-1) -- cycle;
\node at (-0.45,-1.2) {$X$};
\node at (0.2,-1.2) {$W$};
\draw (-0.1,-1.4) -- (-0.1,-1);
\draw (0.5,-1.2) -- (7.2,-1.2);
\draw [fill=blue!10] (6.2,-1) -- (6.2,-1.4) -- (9.8,-1.4) -- (9.8,-1) -- cycle;
\draw (7,-1) -- (7,-1.4);
\draw (8.4,-1) -- (8.4,-1.4);
\draw (9.2,-1) -- (9.2,-1.4);
\node at (6.6, -1.2) {$R_I$};
\node at (7.7,-1.2) {$C$};
\node at (8.8,-1.2) {$R_I$};
\node at (9.5,-1.2) {$W$};
\draw [fill=red!10] (14,-1) -- (9.8,-1) -- (9.8,-1.4) -- (14,-1.4);
\draw (13.8,-1) -- (13.8,-1.4);
\node at (11.8,-1.2) {$P$};

\draw [dashed,line width=2] (9.8,1.2) -- (9.8,-1.8);

\node at (15.8,0) {};

\end{tikzpicture}
      }
      \captionsetup{width=.8\linewidth}%
      \caption{Case 2: $P + 4W \leq (N - 1) \cdot (C + 2R_I + W)$.
        The lock is always taken.}
      \label{fig:schedule-2}
    \end{subfigure}
    \caption{Examples of executions of the coarse-grained algorithm. Blue intervals depict
      critical sections and red intervals depict parallel sections.}
    \label{fig:execution}
  \end{minipage}\hfill
  \begin{minipage}{0.21\textwidth}
    \centering
    \includegraphics[width=\textwidth]{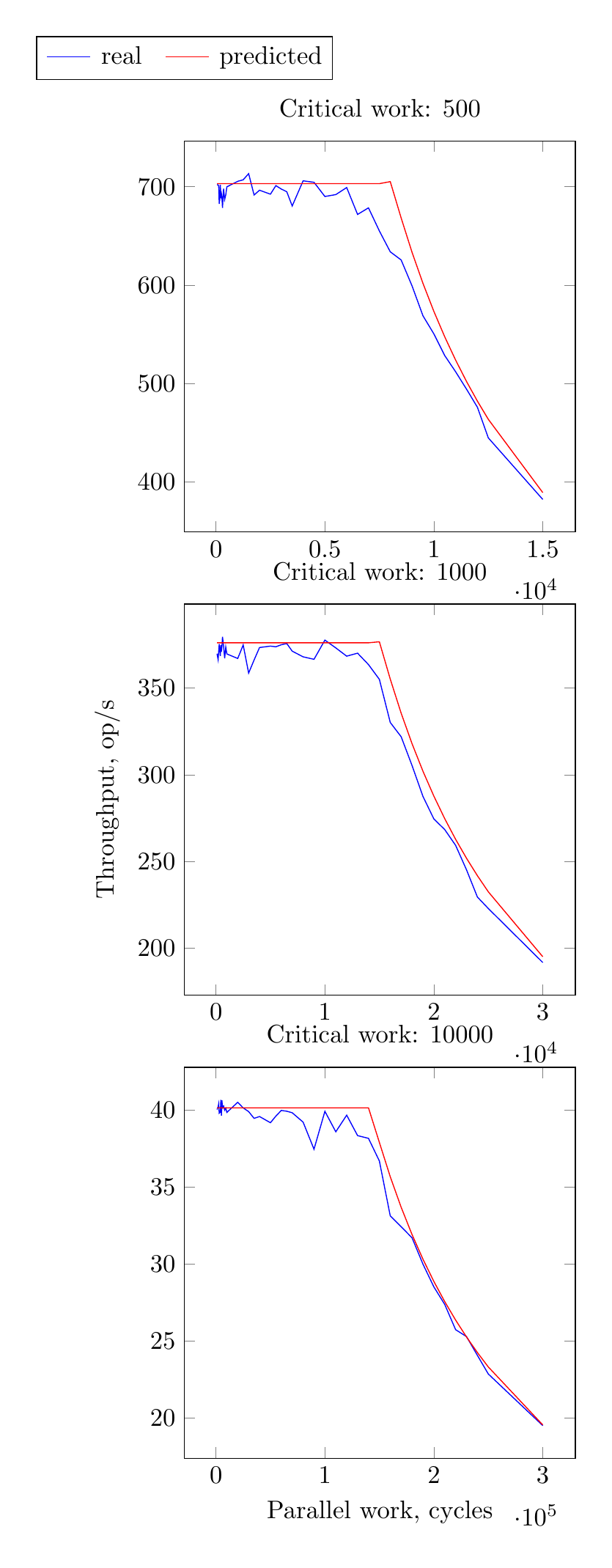}
    \caption{Throughput on $15$ processes for $C \in \{500, 1000, 10000\}$}
    \label{fig:plots}
  \end{minipage}
\end{figure*}

We consider the schedules that correspond to these two cases: the first case is shown in Figure~\ref{fig:schedule-1} and the second is shown in Figure~\ref{fig:schedule-2}.
The complete analysis is presented in Appendix~\ref{sec:analysis}.

\setlength{\belowcaptionskip}{-5pt}
\setlength{\abovecaptionskip}{10pt}

\begin{figure}[H]
\centering
\begin{lstlisting}
class Node:
  bool locked // shared, atomic
  Node next = null 
  
tail = null // shared, global
myNode = null // per process
  
operation():
  myNode.next = null // $W$ |\label{line:setnext}|             
  myNode.locked = true // $W$ |\label{line:setlocked}|                                   |\label{line:start}|
  pred = swap(&tail, myNode) // $W$ or $X$  |\label{line:swap}|
  if pred != null:                                                       |\label{line:pred:condition}|
    pred.next = myNode // $W$                                       |\label{line:queue}|
    while myNode.locked: nop // $R_I$ |\label{line:wait-critical}|
 // CS started   
 for i in 1..C: nop // $C$                      |\label{line:critical}|
 // CS finished  
 if myNode.next == null: // $R_I$ or $R_M$                 |\label{line:release-1}|
   if tail.CAS(myNode,null): // $W$ or $X$         |\label{line:cas}|
     goto parallel
   else:
     while myNode.next == null: nop // $R_I$      |\label{line:wait-node}|
       //pass
 myNode.next.locked = false // $W$       |\label{line:unlocked-next}|
 //Parallel section
 parallel:
 for i in 1..P: nop // $P$                                       |\label{line:parallel}|
\end{lstlisting}
\caption{The coarse-grained operation with inlined \texttt{lock} and \texttt{unlock} functions}
\label{fig:coarse}
\end{figure}

Now the throughput can be expressed as:
%
$\frac{\alpha}{C + 2R_I + W}$ if $P + 4 \cdot W \leq (N - 1) \cdot (C + 2R_I + W)$, and  
$\frac{\alpha \cdot N}{P + C + R_M + 4W}$ if  $P + 2\cdot W \geq (N - 1) \cdot (C + 2R_I + W) + R_I - R_M$.

To describe what is happening in between these two extremes, we can use linear approximation. 


\vspace{-0.4cm}
\section{Experiments}
\vspace{-0.2cm}
The code of our benchmark is written in C++. 
We use an Intel Xeon Gold 6230 machine with 16 cores. 
Based on a single test run, we evaluate the constants as follows: $\alpha \approx 4.04 \cdot 10^5$, $W \approx 15$, $R_I \approx 30$, and $R_M \approx 15$. 
In Figure~\ref{fig:plots}, we depict our measurement results (in blue) and our analytical predictions (in red) for three values of $C \in \{500, 1000, 10000\}$ on $15$ processes. 
Here $x$ axis specifies the size of the parallel section and at the $y$ axis--- the throughput. 
As one can see, our prediction matches the experimental results almost perfectly.
The results of our experiments on an AMD machine are given in Appendix~\ref{sec:amd}.

\vspace{-0.5cm}
\section{Conclusion}
\vspace{-0.2cm}
In this short paper, we showed that in the model presented in~\citep{AAK18-ba} one can accurately predict the throughput of lock-based data structures not only when the lock is CLH, but also when the lock is MCS. Further, we want to close the gap by predicting the performance of data structures with the rest lock-types, such as spin locks, etc. Finally, we want to check whether in this model we can predict the throughput of lock-free data structures, such as Treiber's stack.

\bibliography{references}

\appendix

\section{Analysis}
\label{sec:analysis}

\subsection{The first case. Lock is unlocked.}
We start with the first case. The corresponding schedule is presented on Figure~\ref{fig:schedule-1}. The first process performs two writes each with cost $W$ in Lines~\ref{line:setnext}~and~\ref{line:setlocked}, a contended swap with cost $X$ in Line~\ref{line:swap} while other processes perform two writes with cost $W$, a contended swap with cost $X$, and a write with cost $W$ in Line~\ref{line:queue}. Then, the first process performs: 1)~the critical section with cost $C$; 2)~a read from an invalid cache line with cost $R_I$ in Line~\ref{line:release-1} (on the first try, there is the second process waiting for the lock); 3)~unlock the next process with cost $W$ in Line~\ref{line:unlocked-next}; 4)~the parallel section of cost $P$; 5)~two writes of cost $W$ in Lines~\ref{line:setlocked}~and~\ref{line:setnext}; 6)~a swap with \texttt{null} of cost $W$ in Line~\ref{line:swap} since it is already uncontended; 7)~the critical section with cost $C$; 8)~a read from the exclusive cache line with cost $R_M$ in Line~\ref{line:release-1} since it has not been changed from Line~\ref{line:setnext}; 9)~a CAS of cost $W$ in Line~\ref{line:cas}; 10)~the parallel section of cost $P$; 11)~continue from the point 5) again.

The other processes do the similar thing. All other processes perform: 1)~the invalid read of cost $R_I$ at Line~\ref{line:wait-critical} since there was the previous process waiting for the lock; 2)~the critical section with cost $C$; 3)~an invalid read of cost $R_I$ at Line~\ref{line:release-1} (please, note that process $n$ spends there only $R_M$, since for him the queue is empty); 4)~unlock with cost $W$ in Line~\ref{line:unlocked-next} or, for $n$-th process, make a successful CAS of cost $W$ in Line~\ref{line:cas}; 5)~the parallel section of cost $P$; 6)~continue from the point 5) for the first process.

Please, note that this was possible due to the fact that $R_I \geq W$, otherwise, the second process will not go to the point 6) for the first process and there will be a conflict on \texttt{tail}.
Also, this schedule can happen only when the parallel section with two writes of the first process is bigger than the preliminary work of all other processes: $P + 2W \geq (N - 1) \cdot (2R_I + W + C) + R_M - R_I$. ($R_M$ happens for $n$-th process). And the throughput in that case becomes $\frac{\alpha N}{4W + R_M + C + P}$.

\subsection{The second case. Lock is always taken.}
We continue with the second case. The corresponding schedule is presented on Figure~\ref{fig:schedule-2}. The first process performs two writes each with cost $W$, a contended swap in Line~\ref{line:swap} with cost $X$ while other processes perform two writes with cost $W$, a contended swap with cost $X$, and a write with cost $W$ in Line~\ref{line:queue}.
Then, the first process performs: 1)~the critical section with cost $C$; 2)~a read from an invalid cache line of cost $R_I$ in Line~\ref{line:release-1} since the queue is not empty; 3)~unlock the next process with cost $W$ in Line~\ref{line:unlocked-next}; 4)~the parallel section of cost $P$; 5)~two writes of cost $W$ each in Lines~\ref{line:setlocked}~and~\ref{line:setnext}; 6)~an uncontended swap of cost $W$ in Line~\ref{line:swap}; 7)~adding the process to the queue with cost $W$ in Line~\ref{line:queue}; 8)~wait until the process becomes unlocked with an invalid read with cost $R_I$ in Line~\ref{line:wait-critical}; 9)~continue from the point 1) again.

Any other process start at the point 8) and continue the program of the first process. This case is possible when $P+4W \leq (N - 1) \cdot (2R_I + W + C)$ with the throughput of $\frac{\alpha}{2R_I + W + C}$.

\section{Experiments. Intel.}
\begin{figure*}
    \centering
    \includegraphics[width=\textwidth]{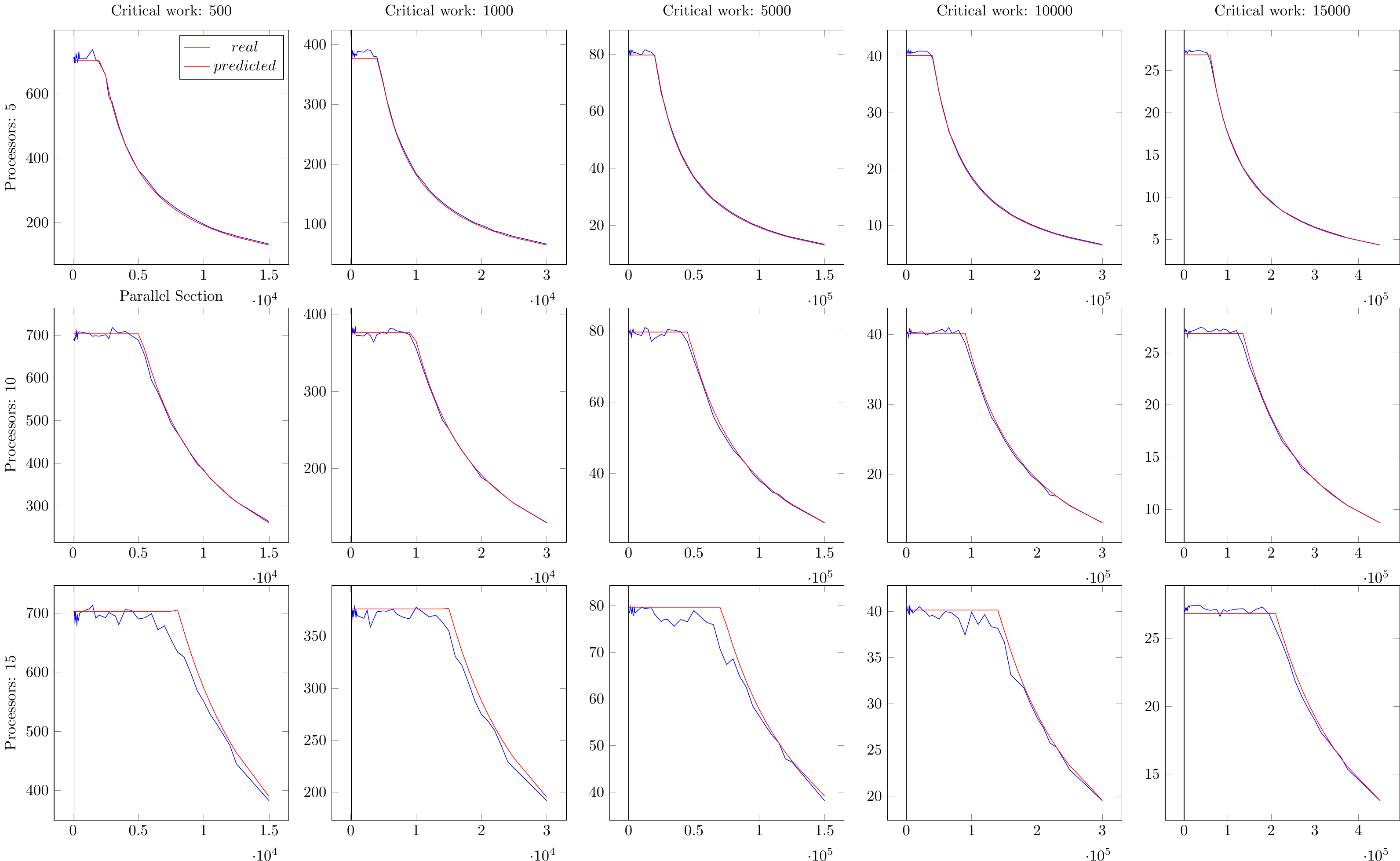}
    \caption{Throughput on $\{5, 10, 15\}$ processes for $C \in \{500,1000,5000,10000,50000\}$. Intel.}
    \label{fig:plots-intel}
\end{figure*}

Results of more experiments on different critical section sizes from $\{500,1000,5000,10000,50000\}$ and the number of processes from $\{5, 10, 15\}$ can be seen on Figure~\ref{fig:plots-intel}.

\section{Experiments. AMD.}
\label{sec:amd}

\begin{figure*}
    \centering
    \includegraphics[width=\textwidth]{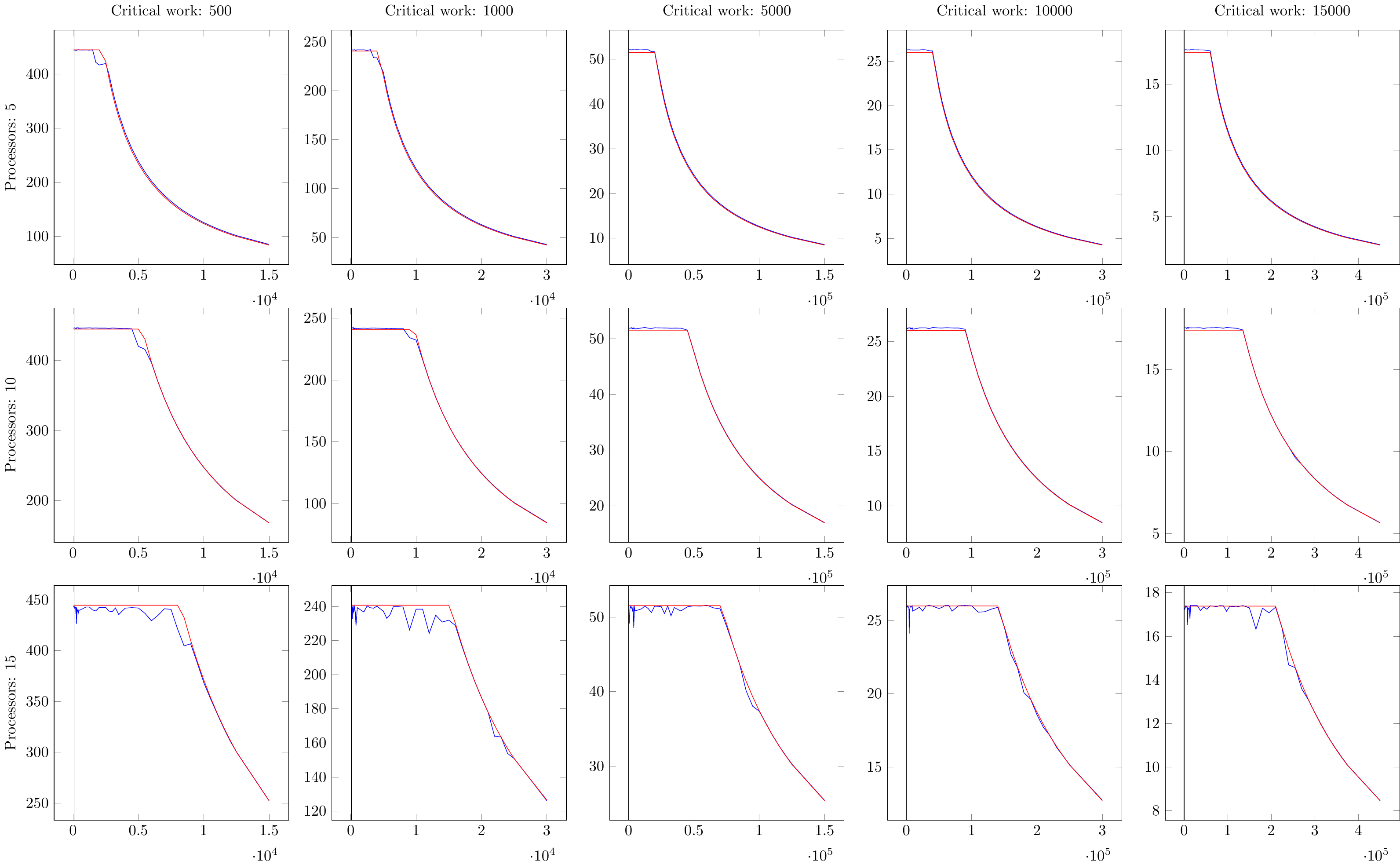}
    \caption{Throughput on $\{5, 10, 15\}$ processes for $C \in \{500,1000,5000,10000,50000\}$. AMD.}
    \label{fig:plots-amd}
\end{figure*}

We use a machine with one AMD  Opteron 6378 with 16 cores. We estimate the constants once during the one chosen run as $\alpha \approx 1.24 \cdot 10^5$, $W \approx 20$, $R_I \approx 35$, and $R_M \approx 15$. The plot~\ref{fig:plots-amd} contains the real-life (blue) line and our expectation (red) line for five different values of critical section $C \in \{500,1000,5000,10000,50000\}$ on $\{5, 10, 15\}$ processes with the size of the parallel section at OX and throughput at OY axis. As one can see our prediction almost matches the experiment.

\end{document}